\documentclass[aps,prc,superscriptaddress,showpacs]{revtex4}
\usepackage[dvips]{graphicx}
\usepackage{amsmath,amssymb}

\newcommand{\Btl}{\mbox{${\tilde B}$}}

\newcommand{\vx}{\mbox{\boldmath $x$}}

\newcommand{\vcr}{\mbox{\boldmath $r$}}

\newcommand{\vlambda}{\mbox{\boldmath $\lambda$}}

\begin{document}

\title{\bf
Spin excitations in fermion condensates}

\author{ Tomoyuki~Maruyama}
\affiliation{
Institute for Nuclear Theory,
University of Washington,
Seattle, Washington 98195, USA}
\affiliation{College of Bioresource Sciences,
Nihon University,
Fujisawa 252-8510, Japan}
\affiliation{Advanced Science Research Center,
Japan Atomic Energy Research Institute,Tokai 319-1195, Japan}

\author{George F. Bertsch}
\affiliation{
Institute for Nuclear Theory,
University of Washington,
Seattle, Washington 98195, USA}

\begin{abstract}
We investigate collective spin excitations in two-component fermion condensates
with special consideration of unequal populations of the two components.
The frequencies of monopole and dipole modes
are calculated using Thomas-Fermi theory and the scaling approximation.
We demonstrate that spin oscillations have more sensitivity to the
interaction and the 
properties of the condensates than the density oscillations.
\end{abstract}


\maketitle

\section{Introduction}

Since the realization of the Bose-Einstein condensed (BEC) atomic 
gases \cite{BECrv,nobel}, 
there has been much interest in ultracold trapped atomic systems 
to study quantum many-body phenomena. 
Besides the Bose-Einstein condensates (BEC) \cite{BECrv,nobel,bec}, 
one can now study degenerate atomic fermi gases \cite{ferG} 
and bose-fermi mixtures \cite{BferM}.
These systems offer great promise to exhibit new and interesting
phenomena of quantum many-particle physics.  An important diagnostic
signal for these systems is the spectrum of collective excitations.
Such oscillations are common to a variety of many-particle systems
and are often sensitive to the interaction and the structure of the
ground state.  
 
However, for harmonically trapped gases of either bosonic or fermionic atoms,
the frequencies of the density oscillations are quite insensitive
to the strength of the interaction.  In particular, the dipole oscillation
frequency is completely independent of the interaction \cite{Kohn}. 
As we will see, excitations where the two components move out of phase
(spin excitations) behave quite differently.

The underlying theoretical tool to treat dynamic problems of
dilute quantum gases is time-dependent mean field theory.  This
reduces to the random-phase approximation (RPA) for small
amplitudes, and the theory in this form has been applied to
density oscillations of these systems \cite{ClFer,Kry}.  When the
single-particle spectrum is regular, the long-wavelength excitations 
are collective and simpler methods can be used to calculate the
frequencies, in particular with sum rules \cite{sumF} 
or the scaling approximation \cite{scal,bohi,berbul}.
In this
work we will follow the last approach, which we believe is justified
by the extreme regularity of the harmonic trapping potential.
The scaling method is physically quite
transparent, but it is in fact equivalent to 
the theory based on energy-weighted
sum rules \cite{bohi}, as used for example in ref.~\cite{sumF}.

In the dilute limit, the interaction in a two-component fermion 
condensate 
is characterized by a single number, the scattering length $a$.  We will
only consider here the case of positive scattering lengths which 
correspond to a repulsive interaction.  When the interaction is
attractive, the system is superfluid in its ground state and the
excitation properties are controlled by the energy gap.

The ground state properties of the two-component fermion system with a 
repulsive interaction was treated by Sogo and Yabu \cite{SY},
who showed that there are three regimes, depending on interaction
strength.  For small interaction strengths, the ground state 
has equal densities of the two components, which we call the
"paramagnetic" regime.  Beyond a certain threshold in interaction strength,
the ground state becomes "ferromagnetic", that is with a unequal
densities of the two components.  At very large interaction strengths,
the minority component vanishes.  We shall call this the single-component
phase.  In this work, we will treat systems in the presence of an
external magnetic field, which will produce a net spin in
the ground state even the paramagnetic regime.

We now briefly describe the spin modes that could be excited by a 
time-dependent external magnetic field.  The simplest field to 
consider is one that is spatially uniform.  
 However, such fields cannot
change the spatial part of the single-particle wave functions
of the condensate and do not induce internal excitations.
The response to a uniform field is identical to that of 
a noninteracting ensemble of $N_1 - N_2$ atoms, where $N_i $ is
the number of atoms in component $i$.  
The next field to consider
has a dipole spatial dependence.  Due to the constraints of
Maxwell's equations, the overall multipolarity of such a field
is $J=2$, but we shall follow the common terminology calling it
dipole.  The spin-dipole mode has been discussed previously in 
in refs.~\cite{sumF} and \cite{ClFer}.  Ref.~\cite{sumF} treats the
the spin dipole mode as overdamped and does not discuss its frequency.
In ref.~\cite{ClFer}, the spin dipole response was calculated in RPA
for a particular value of the coupling strength, and it was found
to be rather narrow and close to the unperturbed oscillation frequency.
In our work here, we will calculate the frequency under a variety
of conditions:  different coupling strengths and in the presence of
an external static magnetic field.  The damping of the mode is beyond
the scope of this study.

\section{Ground state}
\label{grdSec}

We begin with the expression for the energy of a dilute trapped 
condensate in the
mean-field approximation:
\begin{eqnarray}
E_T &=& \int d^3 r ~ [~ - \frac{\hbar^2}{2m}\sum_n^{occ} \sum_{s=1,2}
\psi^{^*}_{n s}    \nabla^2 \psi_{n s}
+ \frac{m}{2} ( \Omega_T^2 r_1^2 + \Omega_T^2 r_2^2 + \Omega_L^2 r_3^2 )
(\rho_1+\rho_2) 
\nonumber \\
& & + g \rho_1\rho_2
- B (\mu \rho_1 -\mu \rho_2)
 ~] .
\label{etot}
\end{eqnarray}
Here $\psi_{n s}$ are orbital wave functions indexed by spin $s=1,2$,
$\Omega_{L,T}$ are the longitudinal and transverse frequencies of the
trapping field, $g$ is the
coupling strength of a contact interaction, $\mu_s$ are the magnetic
moments of the atoms, and $B$ is an external magnetic field.  The densities 
$\rho_s$
are given by the usual sum over occupied orbitals: $\rho_s = \sum_n^{occ} |\psi_{n,s}|^2$. 

We shall use the Thomas-Fermi approximation to evaluate the first term,
the kinetic energy.  We can then make a change 
of variables to simplify the
appearance of the Thomas-Fermi equations, similar to the scaling defined
in 
ref.~\cite{SY}.  With the scaled variables and the Thomas-Fermi
approximation,
the expression for the energy is, up to an additive constant,
\begin{equation}
{\tilde E}_T = \int d^3 x ~ [ \sum_{i=1,2}
\{ \frac{3}{5} n_i^{5/3} + x^2 n_i \} + g n_1 n_2
- {\tilde B} (n_1 - n_2)] .
\label{etotTFs}
\end{equation}
Here the variables are defined
\begin{eqnarray}
x_{j} &=& 
(\frac{m^2 \Omega_j}{ 3 \pi^2 \hbar^3}) r_{j} ~~~~~ (j=1 \sim 3) , 
\nonumber \\
n_i & = &  \left( \frac{m}{\hbar^2} \right)^3 
\frac{2}{9 \pi^4} \rho_i  ~~~~~ (i=1,2)  ,
\nonumber \\
{\tilde B} &=& \left( \frac{m}{\hbar^2} \right)^3 
\frac{2}{9 \pi^4} B \mu, 
\nonumber \\
{\tilde E}_T &=& \frac{4 m^{12} \Omega_L\Omega_T^2}
{(3 \pi^2)^7 \hbar^{21}}  E_T .
\label{etotsc}
\end{eqnarray}
The TF equations for the densities $n_{1,2}$ are derived by variation of
the energy (3) with a constraint on the total number of particles.  
This yields
\begin{eqnarray}
n_1^{2/3} + g n_2 &=& e_f - x^2 + {\tilde B}
\\
n_2^{2/3} + g n_1 &=& e_f - x^2 -{\tilde B}
\label{TFeq}
\end{eqnarray}
In this equation the Lagrange multiplier $e_f$ has the meaning of the 
Fermi energy.
The solution of these equations in the absence of a magnetic field
is discussed in detail in ref.~\cite{SY}.  These authors make an
additional rescaling to eliminate $g$, but we do not do that here.
For any positive $g$, there is a value of $e_f$ above which the
system becomes ferromagnetic.  At $g=1$, the critical point is
at $e_f = 20/27$.  Below that value, the minimum energy is obtained
for the paramagnetic phase, having $n_1=n_2$.  Just above that
value, both components are present but the densities are unequal.
When $e_f$ is increased past $e_f=1$, the system goes
into a single-component phase in the center of the trap, at $x=0$.
Away from the center of the trap, the Fermi energy is effectively
reduced, allowing the paramagnetic phase to persist in the outer
part of the condensate cloud.  Thus the minority component will
form a hollow sphere, which we will call the hollow spin phase.

Now consider the effect of a magnetic field.  
If $\Btl \ne 0$, the densities
$n_{1,2}$ will unequal no matter what the value of $g$.  
Still, one can distinguish two kinds of behavior 
near the center of the trap.  
If the densities of both spins decrease as one moves away from the center,
we shall call it paramagnetic.  
However, for a certain range of parameters,
the density of the minority component may increase initially, moving away
from the center.   
This is the ferromagnetic phase.  
The system can also
form the hollow spin phases in the presence of a magnetic field.  
When the external magnetic field $\Btl$ increases more, 
the minority component disappears completely, giving what we
call the single-spin phase.

\begin{figure}[ht]
\vspace{0.5cm}
\hspace*{0cm}
\includegraphics[scale=0.8]{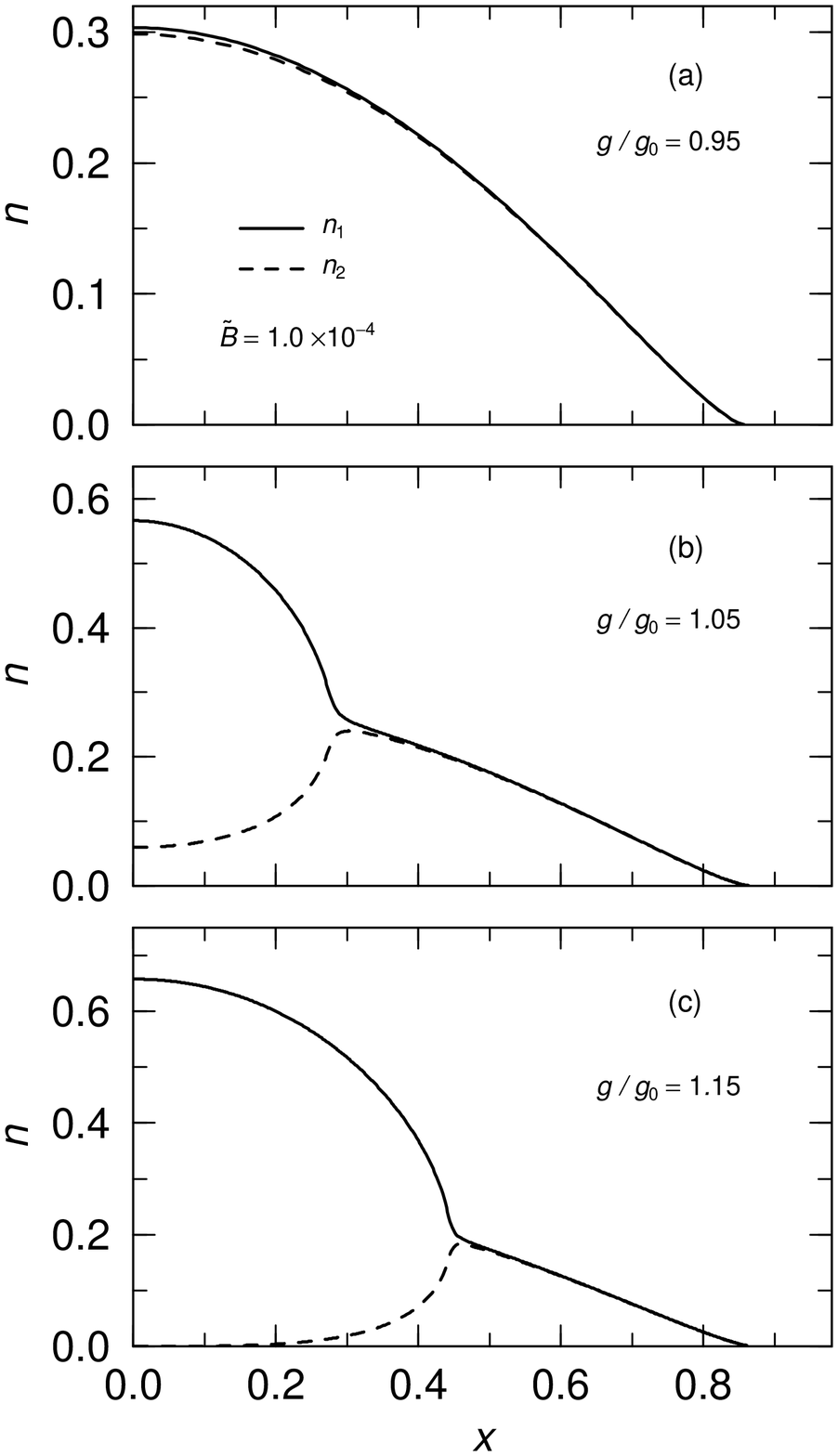}
\caption
{\small 
The scaled density distribution of two components fermion
at $g = 0.95$ (a), $g = 1.05$ (b) and $g = 1.15$ (c)
with with $e_f=20/27$ and $g^2 {\tilde B} = 1.0 \times 10^{-4}$. 
The solid and dashed lines represents the distribution
of the major and minor components, respectively. 
}
\label{frho}
\end{figure}

Density profiles illustrating these three regimes are shown in Fig.
\ref{frho}.  
The panels show the densities as a function of distance $x$
in a weak magnetic field ($g^2 \Btl = 1.0 \times 10^{-4}$)
and at three different interaction strengths:
$g = 0.95$(a), 1.05(b) and 1.15(c).
From top to bottom, the panels show paramagnetic (a), ferromagnetic (b), 
and the hollow spin phases (c).  
The solid and dashed
lines represent the scaled density distribution of major and minor components
of fermions.

\begin{figure}[ht]
\vspace{0.5cm}
\includegraphics[scale=0.85]{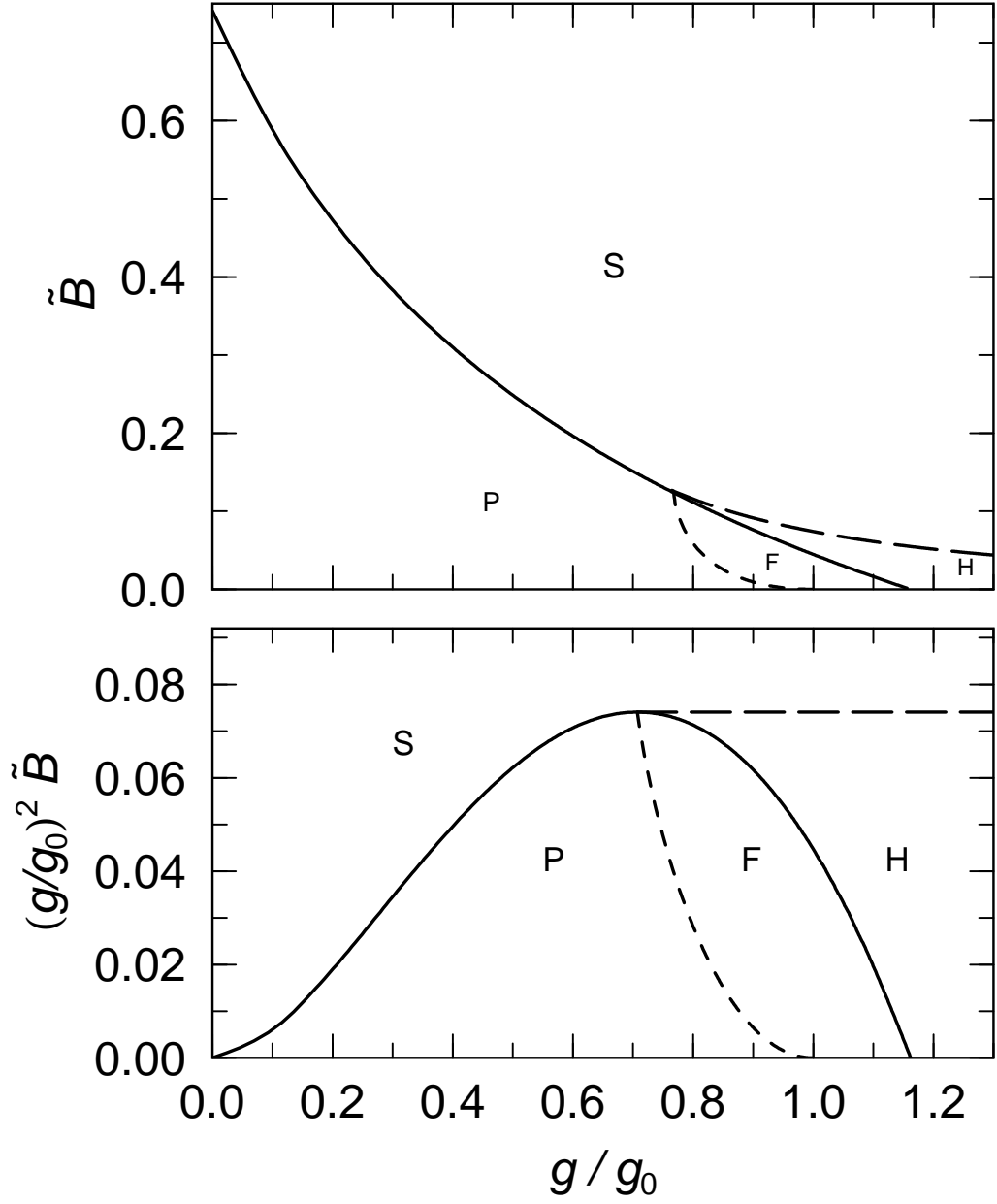}

\caption
{\small 
The phase diagram of the two components of fermion.
The symbols ''S'', ''P'', ''F'' and ''H'' denote
the region of the single-component, paramagnetic, ferromagnetic, single 
and hollow phases.}
\label{phdg}
\end{figure}

The phase boundaries as function of $\Btl$ and
$g$ are shown in Fig. \ref{phdg}.
The upper and lower columns show same results, but
in the lower column the vertical line is rescaled by
factor $g^2$. 

The dashed lines represents the border between the paramagnetic and
ferromagnetic phases, which is given by
 $\partial^2 n_2(x)/ \partial x^2 = 0$ at $x=0$.

The solid line represents the results
solved by $n_2(0) =0$ in eq.(\ref{TFeq}).
This line crosses the dashed line at $g = \sqrt{2}$ and $\Btl=1/27$.
The curvature of the minority component density, 
$\partial^2 n_2(x)/ \partial x^2$ at $x=0$, is negative
when $g < 1/\sqrt{2}$ and positive when $g > 1/\sqrt{2}$
This solid line shows the border between the single and paramagnetic spin 
phases when $g < 1/\sqrt{2}$, and
the border between the ferromagnetic and hollow spin phases
when $g < 1/\sqrt{2}$.

In the hollow phases
there are two boundaries for the minority fermion density  
where $n_2 = 0$ and $n_1 \neq 0$.
On this boundary the density of majority fermion must satisfy
\begin{equation}
n_1^{\frac{2}{3}} - g n_1 = 2 {\tilde B} .
\end{equation}
Here we should consider the following function:
\begin{equation}
f(n) = n^{\frac{2}{3}} - g n - 2 {\tilde B} .
\end{equation}
The derivative of the above equation is
\begin{equation}
\frac{d}{d n} f(n) = f^{\prime} (n) = \frac{2}{3} n^{-\frac{1}{3}} - g .
\end{equation}
Since $f^{\prime}(8/27g^3)=0$,
$f(0) =  f(1/g^3) = - 2B < 0$ and $f^{\prime}(1/g^3)<0$, $f(n)$
becomes maximum at $n=8/27 g^3$. 
Thus $f(8/27g^3)>0$ is required, otherwise $f(n) \le 0$, and there is no boundary of $n_2$,
which means $n_2=0$ in all regions.
From that we can obtain the condition for ${\tilde B}$ as
\begin{equation}
{\tilde B} < {\tilde B}_{max} 
= \frac{1}{2} \left( \frac{4}{9g^2} - \frac{8}{27g^2} \right)
= \frac{2}{27g^2} \approx 0.074 g^{-2}.
\end{equation}  
The long-dashed represents $\Btl= 2/27g^{-2} \approx 0.074 g^{-2}$
for $g > 1 / \sqrt{2}$,
which means the border between the hollow and single
spin phases.

\begin{figure}[ht]
\vspace*{1.0cm}
%\hspace*{0.5cm}
\includegraphics[angle=270,scale=0.6]{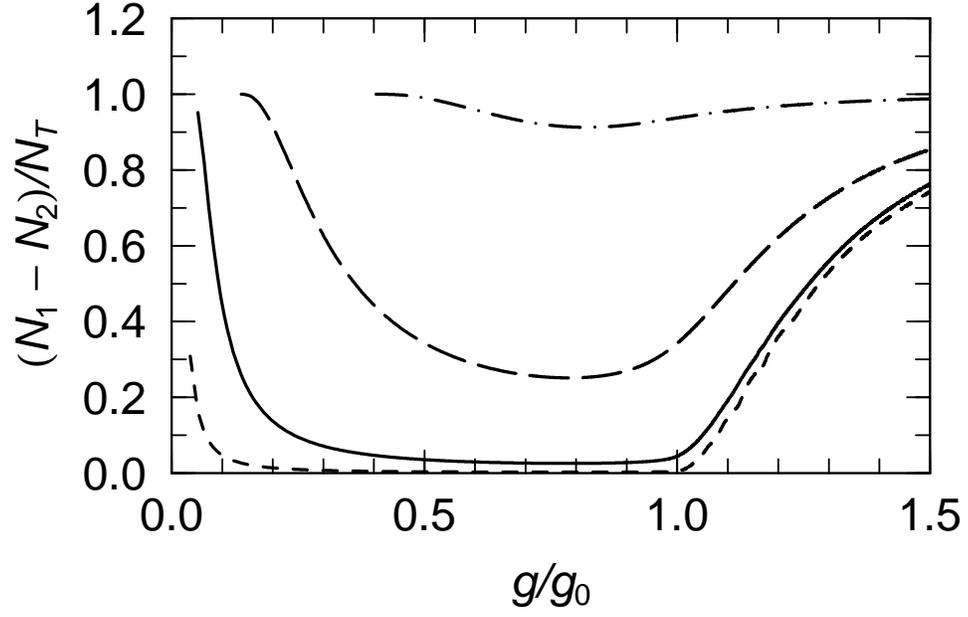}

\bigskip
\caption
{\small 
The asymmetry of the two components 
 as a function of the coupling constant $g$., at $e_f = 20/27$. 
The dashed, sold. long-dashed and dotted lines represent
the results with the external magnetic fields 
$g^2 {\tilde B} =$ $1.0 \times 10^{-4}$,  $1.0 \times 10^{-3}$, 
 $1.0 \times 10^{-2}$ and  $5.0 \times 10^{-2}$,
respectively.}
\label{nasym}
\end{figure}

In Fig.~\ref{nasym} we plot 
the number asymmetry between two components $(N_1 - N_2)/N_T$,
where $N_T = N_1 + N_2$. 
When the external magnetic field is small,
the asymmetry of the fermion number is very small 
and almost independent of the coupling in the paramagnetic phase
($g \lesssim 1.0$) except when the coupling is very small,
where the system is close to the border 
between the paramagnetic and single spin phases. 
As the coupling increases further and 
exceeds about $g \approx 1.0$, the system is
changed from the paramagnetic spin phase to 
the ferromagnetic one.
The number asymmetry of the fermion number also increases 
As the external magnetic field becomes larger,
the change of the number asymmetry is not drastic,
though the rapid change of the dipole frequency still remains.

\section{Dipole excitations}

In this section we derive an expression for the dipole frequency
using the scaling method. Without loss of generality, we take the
direction of motion along the $z$-axis.  The time-varying density
will be parameterized with a collective coordinate $\lambda_i$
as \cite{scal}.
\begin{equation}
\rho_i (\vcr) = \rho_i^{(0)}(r_1, r_2, r_3 - \lambda_i) ,
\label{dscl}
\end{equation}
where $\rho_i^{(0)}(\vcr)$ is the density distribution of the ground state.
Substituting eq.(\ref{dscl}) into eq. (\ref{etot}), we can obtain the
variation of the total energy up to the order of $O(\lambda^2$) as
\begin{eqnarray}
\Delta E_T &=& E_T - E^{(0)}_T 
\nonumber \\
&\approx& 
- \frac{1}{2} m \Omega_L^2 (\lambda_1^2 N_1 + \lambda_2^2 N_2 ) +
A (2 \lambda_1 \lambda_2  -  \lambda_1^2  - \lambda_1^2 ), 
\end{eqnarray}
where ${\tilde E}^{(0)}_T$ is the ground state energy ($\lambda_i = 0$), 
and 
\begin{equation}
 A = \frac{g}{2} \int d^3 r ~ 
\frac{\partial \rho^{(0)}_1}{\partial z} 
\frac{\partial \rho^{(0)}_2}{\partial z} .
\end{equation}

When we consider the time-dependence of $\lambda_i$,
the mass parameter with respect to $\lambda_i$ is obtained as 
$m N_i$ \cite{scal}.

Then the classical equation of motion for $\vlambda$ is harmonic, 
giving rise to the following eigenvalue equation for the oscillation
frequencies,
\begin{eqnarray}
\left( \begin{array}{cc} m(\Omega_L^2-\omega^2) N_1 -  2 A &  2 A \\
 2 A &  m (\Omega_L^2-\omega^2)N_2 -  2 A \end{array} \right)
\left(\begin{array}{c} \lambda_1 \\ \lambda_2 \end{array} \right) = 0
\end{eqnarray}

The eigenvalues of this equation are given\begin{equation}
\omega^2 = \Omega_L^2 - ( \frac{A}{mN_1}+ \frac{A}{mN_2})
\pm  ( \frac{A}{mN_1}+ \frac{A}{mN_2}) .
\end{equation}

One eigenvalue is  $\omega= \Omega_L $ with the eigenvector
$\lambda_1 = \lambda_2$. 
This is the in-phase oscillation of the system which follows
Kohn's theorem.

The other eigenvalue is 
\begin{eqnarray}
\omega  &=& \omega_D \equiv \{ \Omega_L^2- 
( \frac{2 A}{mN_1} + \frac{2 A}{N_2} ) \}^{\frac{1}{2}} .
\label{frDP}
\end{eqnarray}
The eigenvector is given by 
$\lambda_1/N_1 = -\lambda_2/N_2$.
In this mode, the two components move in opposite directions
keeping  their center of mass
at rest.

Here we should give a comment.
If we consider the dipole oscillation in the transverse relation,
the frequency normalized by the transverse trapped frequency $\omega_D/\Omega_T$
is also equivalent to the right-hand-side of eq.(\ref{frDP}).

Using the variable transformation explained 
in the previous sections,  
the dipole frequency is written as
\begin{eqnarray}
\omega_D/ \Omega_{L(T)}  &=& \{ 1 - 
{\tilde A}
( \frac{1}{{\tilde N}_1} + \frac{1}{{\tilde N}_2} ) \}^{\frac{1}{2}}, 
\end{eqnarray}
where
\begin{eqnarray}
{\tilde A} &=& \frac{g}{6} \int d^3 x ~ 
\frac{\partial n_1}{\partial x} \frac{\partial n_2}{\partial x} ,
\label{rsfA}
\\
{\tilde N}_i &=& \int d^3 x ~  n_i
=  \frac{2 m^9} {3^5 \pi^{10} \hbar^{15}}  N_i .
\label{tlN}
\end{eqnarray}
 
\begin{figure}[ht]
\vspace{0.5cm}
\hspace*{0.5cm}
\includegraphics[scale=0.6,angle=270]{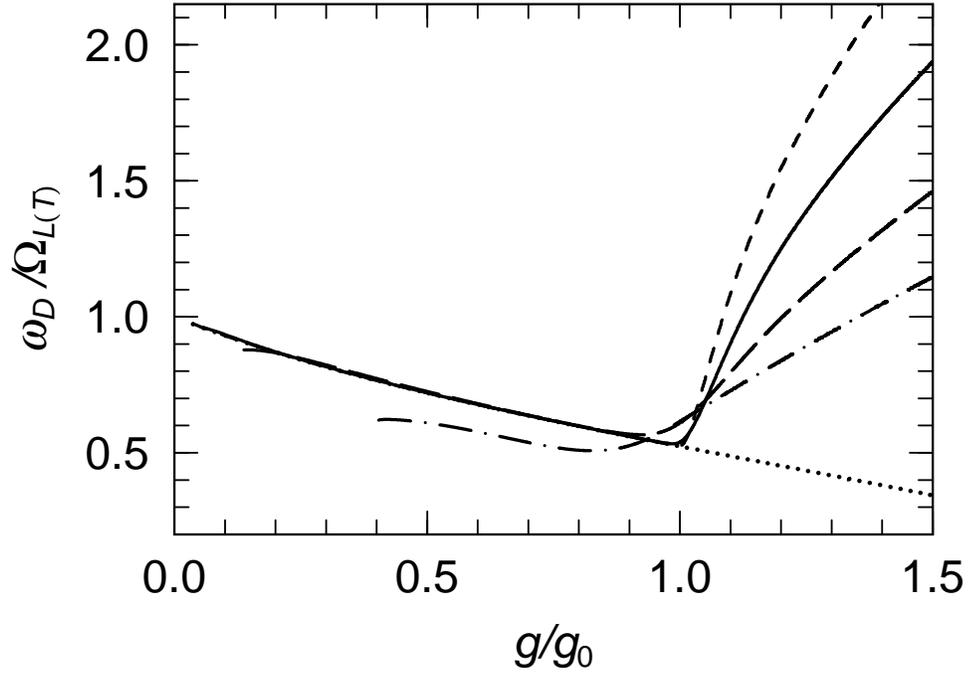}

\caption
{\small 
The frequencies of the dipole oscillations
 as a function of the coupling constant $g$.
The dashed, sold. long-dashed and dotted lines represent
the results with the external magnetic fields 
$g^2 {\tilde B} =$ $1.0 \times 10^{-4}$,  $1.0 \times 10^{-3}$, 
 $1.0 \times 10^{-2}$ and  $5.0 \times 10^{-2}$,
respectively.
The dotted line denote the result of the symmetric system.}
\label{dqplg}
\end{figure}

In Fig.~\ref{dqplg} we show the frequency of the dipole oscillation 
in  the out-of-phase oscillation
as a function of the coupling constant, $g$, 
with various external magnetic fields, 
$g^2 {\tilde B} = 1.0\times 10^{-4}$ 
(dashed), $1.0 \times 10^{-3}$ (solid), 1.0$\times 10^{-2}$ (long-dashed), and  
5.0$\times 10^{-2}$ (chain-dotted), respectively.
For comparison we also show the frequencies in the symmetric system
in zero magnetic field (dotted line).
 
As the coupling becomes larger,
the dipole frequency decreases monotonically
until $g \approx 1.0$, and sharply increases above that.
For $g \lesssim 1.0$, where the number asymmetry is small,
the frequency $\omega_D$ does not strongly depend on 
the external magnetic field.
In the paramagnetic spin phase ($g \lesssim 1$), 
the density distribution of the minor component of fermion $n_2$ is similar to 
that of the major component $n_1$, and the integrand 
in ${\tilde{A}} > 0$ (\ref{rsfA})
is positive in all regions.
As the coupling increases, the size of the fermi gas becomes larger, and then
the frequency $\omega_D$ monotonously decreases.

As the coupling becomes  $g \gtrsim 1$ 
one component of the fermions is partially converted into the other component, 
and ferromagnetism appears in the central region.
In Fig.~\ref{frho}b
$\partial n_2/\partial x > 0$ for $x < x_c$ and 
$\partial n_2/\partial x < 0$ for $x > x_c$, where $x_c \approx 0.32$;
namely there is a ferromagnetic region for $x < x_c$. 
The contribution from this ferromagnetic region to ${\tilde{A}}$
 is negative.
As the coupling increase, the critical position $x_c$ moves to the surface,
${\tilde A}$ 
 becomes smaller, and then $\omega_D$  increases.

In the case of a strong magnetic field, the qualitative
behavior is similar.
In strong coupling  $g \gtrsim 1$ the slope of the density function 
for the minor fermion $n_2$ is smaller in the strong magnetic field 
than that in the weak magnetic field.
As the external magnetic field increases, 
the dipole frequency becomes 
smaller.

\newpage

\section{Monopole Oscillation}

In this section we study the monopole oscillation in spherical trap,
$\Omega_T=\Omega_L=\Omega_M$.
Generally the monopole and quadrupole oscillations are coupled,
but qualitative properties are not so significant difference
between the two collective oscillations.
In order to study the monopole oscillation,
we introduce the following scaling:
\begin{equation}
\rho_i(\vcr) = e^{3 \lambda_i} \rho^{(0)}_i (e^{\lambda_i} \vcr).
\end{equation} 
Under this scaling the total energy becomes
\begin{equation}
E_T = \sum_{i=1,2} \{ e^{2\lambda_i} T_i 
+ e^{-2\lambda_i} U_{i} \}
+ V_{12} ,
\end{equation}
where the $T_i$ and $U_{i}$ are the kinetic and harmonic 
oscillator energy parts in the ground state, respectively.
The interaction energy $V_{12}$ appears
\begin{eqnarray}
V_{12} &=& g e^{3\lambda_1 + 3 \lambda_2} 
 \int d^3 r ~ \rho^{(0)}_1(e^{\lambda_1}\vcr)  \rho^{(0)}_2(e^{\lambda_2}\vcr) 
\nonumber \\
&\approx&
- g \int d^3 r ~  \{ \lambda_1 \rho^{(0)}_1 \vcr\frac{\partial \rho^{(0)}_2}{\partial \vcr}
+ \lambda_2  \vcr \frac{\partial \rho^{(0)}_1}{\partial \vcr} \rho^{(0)}_2 \}
+
 K_{11} \lambda_1^2 +  K_{22} \lambda_2^2
+ 2 K_{12} \lambda_1 \lambda_2
\end{eqnarray}
with
\begin{eqnarray}
K_{11} &=& - \frac{g}{2}  \int d^3 r ~ \{ (3 +  \vcr \frac{\partial}{\partial \vcr}) \rho^{(0)}_1 \}
\{ \vcr \frac{\partial}{\partial \vcr} \rho^{(0)}_2\} ,
\\
K_{22} &=& - \frac{g}{2} \int d^3 r ~ \{ \vcr \frac{\partial }{\partial \vcr} \rho^{(0)}_1\}
\{ (3 +  \vcr \frac{\partial}{\partial \vcr}) \rho^{(0)}_2 \} ,
\\
K_{12} &=&
\frac{g}{2}  \int d^3 r ~
\{ (3 +  \vcr \frac{\partial}{\partial \vcr}) \rho^{(0)}_1 \}
\{ (3 +  \vcr \frac{\partial}{\partial \vcr}) \rho^{(0)}_2 \} .
\end{eqnarray}

The mass parameter of the monopole oscillation is given by 
the mass times the mean square radius.
Then the monopole oscillation mode and frequencies $\omega_M$ are obtained by
\begin{equation}
\left[ B_M \omega^2 - C_M  \right] \vlambda  = 0 .
\end{equation}
with 
\begin{eqnarray}
B_M &=& \frac{m}{2} \left( \begin{array}{cc} X_1 & 0 \\
 0 & X_2 \end{array} \right) ,
\\
{\cal C}_M &=& 
\left( \begin{array}{cc} C_{11}   & K_{12} \\
 K_{12} &   C_{22}  \end{array} \right) 
=
\left( \begin{array}{cc} 2 T_1 + 2 U_1 + K_{11}   & K_{12} \\
 K_{12} &   2 T_2 + 2 U_2 + K_{22}  \end{array} \right) ,
\end{eqnarray}
where
\begin{eqnarray}
X_i &=&
\int d^3 r \rho_i (\vcr) \vcr^2 .
\end{eqnarray}

Using the variable transformation explained 
in the Sec.~\ref{grdSec},  
\begin{equation}
\left\{ (\frac{\omega_M}{\Omega_M})^2{\tilde B}_M
 - {\tilde C}_M  \right\} \vlambda = 0
\label{Monoeq}
\end{equation}
with
\begin{eqnarray}
{\tilde B}_M &=& \left( \begin{array}{cc} {\tilde X}_1 & 0 \\
 0 & {\tilde X}_2 \end{array} \right) ,
\\
{\tilde C}_M &=& \left( \begin{array}{cc} 2 {\tilde T}_1 + 2 {\tilde X}_1 + {\tilde K}_{11}   
& {\tilde K}_{12} \\
{\tilde K}_{12} &   2 {\tilde T}_2 + 2 {\tilde X}_2 + {\tilde K}_{22}  \end{array} \right) ,
\end{eqnarray}
where
\begin{eqnarray}
{\tilde X}_i &=& \int d^3 x ~ n_i (\vx) \vx^2
\\
{\tilde T}_i &=& \frac{3}{5} \int d^3 x ~ [n_i  (\vx)]^{\frac{5}{3}} ,
\\
{\tilde K}_{11} &=& - \frac{g}{2}
\int d^3 x ~ ( 3 n_1 +  x\frac{\partial n_1}{\partial x}) 
( x \frac{\partial n_2}{\partial x} )  ,
\\
{\tilde K}_{22} &=& - \frac{g}{2} \int d^3 x ~ ( x \frac{\partial n_1}{\partial x} )
 (3n_2 +  x \frac{\partial n_2}{\partial x} ) ,
\\
{\tilde K}_{12} &=& \frac{g}{2}
\int d^3 x ~
 (3 n_1 +  x\frac{\partial n_1}{\partial x})
 (3 n_2 +  x\frac{\partial n_2}{\partial x})
= \frac{g}{2} \int d^3 x ~
\left( x^2 \frac{\partial n_1}{\partial x} 
\frac{\partial n_2}{\partial x} \right) .
\end{eqnarray}

In the symmetric system $\rho_1 = \rho_2$ the monopole frequencies are given by
$\omega_M^2 = ( C_{11} + K_{12}) / m X_1$ and 
$( C_{11} - K_{12}) / m X_1$.
The eigenvector becomes $\lambda_1 = \lambda_2$ (in-phase) in the respect to
the former frequency, and  $\lambda_1 = -\lambda_2$ (out-of-phase) 
in the latter frequency.

\begin{figure}[ht]
\vspace{0.5cm}
\hspace*{0cm}
\includegraphics[scale=0.8]{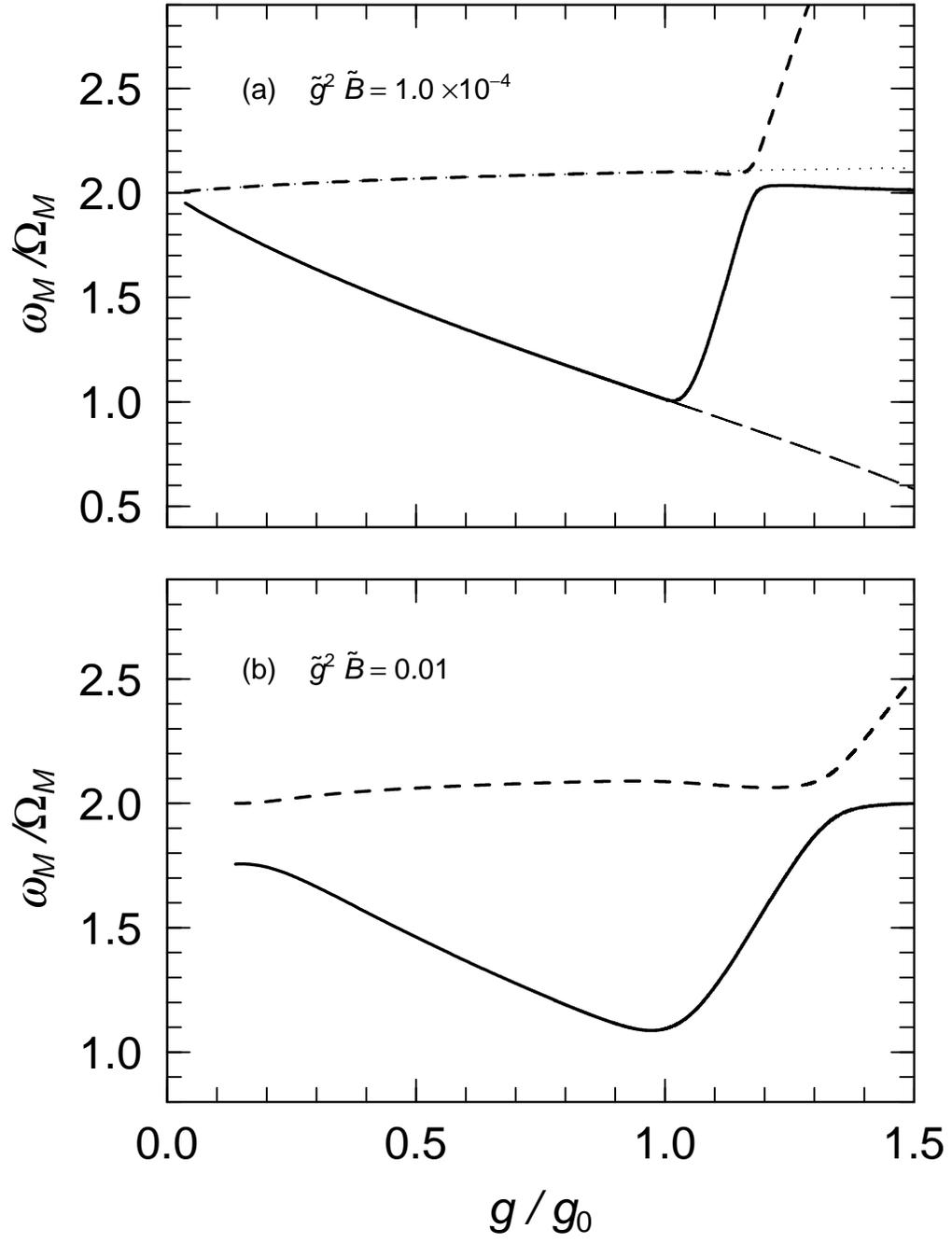}
\caption
{\small 
The frequency of the monopole oscillations.
with the external magnetic fields with 
$g^2 {\tilde B} =$ $1.0 \times 10^{-4}$ (a) and
 $1.0 \times 10^{-2}$ (b).
The thick solid and dashed lines represent the results of
the out-of-phase and in-phase oscillation modes, respectively. 
The thin solid and dashed lines denote the result 
the out-of-phase and in-phase oscillation modes
of the symmetric system, respectively. }
\label{mplg}
\end{figure}

In Fig.~\ref{mplg} we show the frequencies of monopole oscillations 
as functions of the coupling constant $g$ with 
$g^2 {\tilde B} = 1.0 \times 10^{-4}$ (a) and 
$1.0 \times 10^{-2}$ (b).
The solid and dashed lines represent the smaller and larger frequencies
of two modes, which we call  mode-1 and mode-2, respectively.

In Fig.~\ref{mplg}a we also plot
the frequencies of out-of-phase and in-phase modes in the symmetric system 
without external magnetic field
with the thin long-dashed and dotted lines, respectively.
The frequencies of the in-phase oscillation in the symmetric system
is  about $\omega_M \approx 2 \Omega_M$, which is the frequency 
in a non-interacting trapped ideal gas, for any asymmetry.
As the coupling constant increases, on the other hand, the frequency of the out-of-phase mode
monotonously decreases.

In the weak coupling $g \lesssim 1$, 
the frequencies of the mode-1 and mode-2 are 
almost same as those of the in-phase and out-of-phase modes in the symmetric system.
As the coupling constant increases, on the other hand, the frequency of the mode-2 
rapidly increases above $g \approx 1$.
When the coupling constant increases further,
the frequency of the out-of-phase mode approaches to that of the in-phase mode,
and the level mixing between the two modes occurs.
Even after the level mixing, the frequency of the  mode-1
increases while the frequency of the mode-2 rarely vary.
From those we can suppose that the modes-1 and -2
are almost out-of-phase and in-phase modes, 
though the two modes are mixed and exchange their roles in strong coupling.

In order to confirm that, here, 
we calculate the mixing angle $\theta_m$ defined as 
\begin{equation}
{\vlambda} = 
\left( \begin{array}{c} \lambda_1 \\ \lambda_2 \end{array} \right)
= \frac{1}{\sqrt{2}}
\left( \begin{array}{c} 1 \\ 1 \end{array} \right) \cos \theta_m
+
\frac{1}{\sqrt{2}}
\left( \begin{array}{c} -1 \\ ~1 \end{array} \right) \sin \theta_m ,
\end{equation}
where ${\vlambda}$ is an eigenvector obtained from eq.(\ref{Monoeq}).
When $\theta_m = 0, \pi$ and  $\theta_m = \pi/2$,
the modes exhibit the pure in-phase and out-of-phase
 monopole oscillation of two-component fermions.

\begin{figure}[ht]
\vspace{0.5cm}
{\includegraphics[scale=0.85]{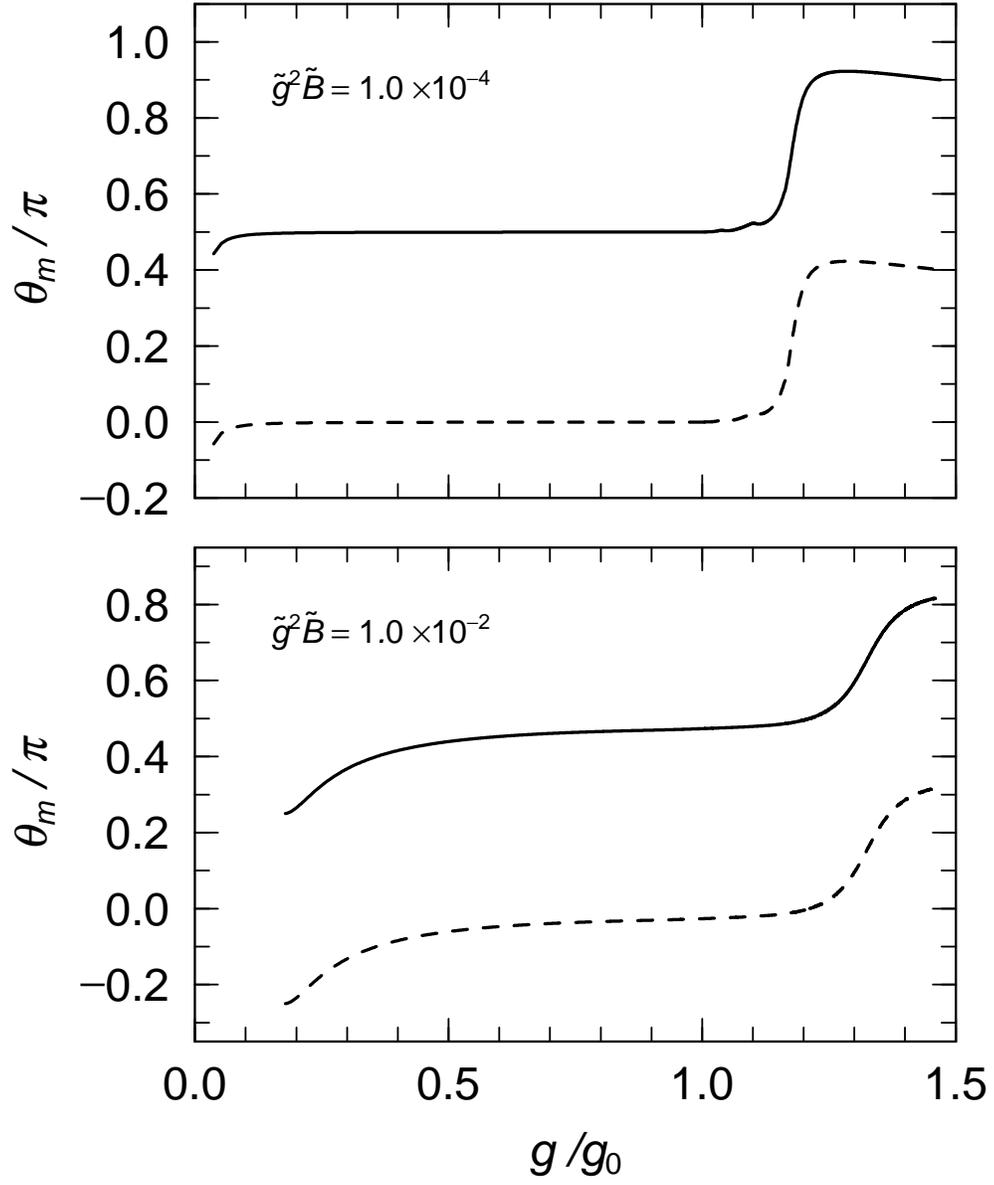}}

\caption
{\small 
The mixing angle of the two modes of the monopole oscillation
with $g^2 {\tilde B} =$ $1.0 \times 10^{-4}$ (a) and
 $1.0 \times 10^{-2}$ (b).
The thick solid and dashed lines represent the results of
the mode-1 and mode-2 , respectively.
 }.
\label{mxang}
\end{figure}

In Fig.~\ref{mxang} we show the results with the external magnetic
fields $g^2 {\tilde B} = 1.0 \times 10^{-4}$ (a) and  
$g^2 {\tilde B} = 1.0 \times 10^{-2}$ (b).
In the case of the small magnetic field (Fig.~\ref{mxang}a), 
the two modes, the mode-1 and the mode-2, are almost clear out-of-phase
and in-phase oscillations before the level mixing.
Around $g \approx 1.2$, the modes-1 and -2 are mixed, and
the two collective modes exchange their roles in the strong interaction limit.
This feature is less clear, but also seen
in the case of the strong magnetic field (Fig.~\ref{mxang}b).

\section{Summary}

In this paper we study the spin excitation on the dipole and
monopole oscillations in the asymmetric two-component fermion condensed
system using the scaling method.
As for the dipole oscillations the in-phase and the
out-of-phase modes are completely decoupled even in the asymmetric
system.
The two modes can be coupled in the monopole oscillation, but these
two modes are rarely mixed except when their frequencies 
are close.  

In all kinds of oscillations the in-phase motions do not
have any special behavior even if the two components are largely asymmetric.
The frequencies of the in-phase modes are not so different from
those of the non-interacting system.

On the other hand the frequencies of these out-of-phase oscillations
monotonously decreases as the repulsive force 
between the two kinds of fermions becomes larger.
As the coupling further increases,
the system changes to ferromagnetic and,
the frequencies of the out-of-phase oscillations 
very suddenly increase,
thought those of the in-phase oscillations do not vary even if the ferromagnetism appears.
In addition the frequencies and M1 excitation strength show 
differences between different number asymmetry for $g \gtrsim 1$, 
where the ferromagnetism appears.

Thus significant information of many body systems can be gotten only from the out-of-phase oscillations.
In this work we describe the two components of fermions with different two spin states.
It may be thought to be difficult to establish such experiments because we need two kinds of 
external magnetic fields to control the interaction and the asymmetry.
However a similar work can be done in systems including two different kinds of fermi gases,
where we do not need an external magnetic field to construct the asymmetry.
If we use two kinds of fermions with different masses, we can study a variety of systems.

\bigskip
{\bf Acknowledgement}

T.M. thank the Institute for Nuclear Theory at  University of Washington for the hospitality
and partially support during the completion of this work.

\newpage

\newpage

\newpage

\newpage 

\newpage

\newpage

\end{document}